\documentclass[twocolumn,reprint,aip,fleqn]{revtex4-1}
\usepackage[fleqn]{amsmath}
\usepackage{amssymb}

\newcommand{\vdot}{\vec{\cdot}}
\newcommand{\rv}{\vec{r}}
\newcommand{\ru}{\vec{\hat{r}}}
\newcommand{\vv}{\vec{v}}
\newcommand{\vu}{\vec{\hat{v}}}
\newcommand{\av}{\vec{a}}

\newcommand{\adot}{\dot{a}}
\newcommand{\avdot}{\vec{\dot{a}}}
\newcommand{\avdddot}{\vec{\dddot{a}}}
\newcommand{\Av}{\vec{A}}
\newcommand{\Ev}{\vec{E}}

\newcommand{\Hv}{\vec{H}}

\newcommand{\vcross}{\vec{\times}}
\newcommand{\vnabla}{\vec\nabla}

\newcommand{\rones}{\rv_\mathrm{1S}}
\newcommand{\rtwos}{\rv_\mathrm{2S}}
\newcommand{\rthrees}{\rv_\mathrm{3S}}
\newcommand{\tones}{t_\mathrm{1S}}
\newcommand{\ttwos}{t_\mathrm{2S}}
\newcommand{\tthrees}{t_\mathrm{3S}}

\newcommand{\ronef}{\rv_\mathrm{1F}}
\newcommand{\rtwof}{\rv_\mathrm{2F}}
\newcommand{\rthreef}{\rv_\mathrm{3F}}
\newcommand{\tonef}{t_\mathrm{1F}}
\newcommand{\ttwof}{t_\mathrm{2F}}
\newcommand{\tthreef}{t_\mathrm{3F}}

\newcommand{\dtfone}{dt_\mathrm{F1}}
\newcommand{\dtftwo}{dt_\mathrm{F2}}
\newcommand{\dtsone}{dt_\mathrm{S1}}
\newcommand{\dtstwo}{dt_\mathrm{S2}}

\newcommand{\psiLW}{\psi_\mathrm{LW}}

\newcommand{\ts}{t_\mathrm{S}}
\newcommand{\mathspace}{\vspace{-20 pt}}
\newcommand{\moremathspace}{\vspace{-15 pt}}

\renewcommand{\vec}[1]{\mbox{\boldmath$#1$}}

\begin{document}

\title{A low order extension the Li\'{e}nard--Wiechert retardation equations
to include the Thomas precession}

\author{G. R. Osborn}
\altaffiliation[]{Anaheim California, USA, Retired}
\email{gary@s-4.com}
\date{\today}

\begin{abstract}

In a calculation that directly parallels the derivation of the Thomas
precession, the first time derivative of the retarded potentials is
derived.  The solutions have to be integrated in time to obtain the
potential solution.

The Thomas precession vanishes when the acceleration and velocity
vectors are parallel, causing the solution for the dipole antenna to be
the same as for the Li\'{e}nard--Wiechert solution, and those solutions
are in turn always solutions to the Maxwell equations.  The solution for
the current loop antenna is not a solution to the Maxwell equations.
Field equations are obtained by restructuring the Proca equations that
are commensurate with the low order retardation solutions.  The
solutions are not in the Lorentz gauge and they are not solutions to the
unmodified Proca equations.

The high order terms are not solutions to the equations.  In
representing angular relationships, an argument is developed that
derivatives beyond the first will be required for more complete
solutions.  The calculations are not in tensor form, but the tensors
represent angular relationships, and the inference is based on the
tensor irreducibility theorem.  In being linear equations expressing
angular relationships, the theorem implies that exact retardation
equations do not exist unless the contravariant tensor of rank $n+1$ is
reducible.
\end{abstract}

\maketitle

\setlength{\mathindent}{8 pt}

\section{Introduction}

If a Lorentz transform is performed to the frame of reference of an
accelerated particle at time $t$, followed by an infinitesimal transform
to the velocity of the particle at time $t+dt$, the result is the same
as a direct transform at time $t+dt$, followed by a space rotation
\cite{aharoni,jackson}.  The
Lorentz transform is a vector equation.  A Lorentz transform followed by
a space rotation is not.  In not being representable with conventional
vector relationships, the space is not flat.

The basis of the precession is that the Lorentz transform does not
achieve closure after transforming full circle through three frames of
reference.  The coordinates in the last frame of reference appear to be
rotated, or rotating if multiple infinitesimal transforms are performed
in the frame of reference of a particle in a circular orbit.

One way of interpreting the relationships is that we should adopt the
conclusions of the observer in the other frame of reference as our own
-- to see ourselves as others see us.  That is because the other frame
of reference could be our frame of reference next time.  The perspective
would not be acceptable if the coordinates in our frame of reference
were spinning, but the retarded potentials are first-known in the frame
of reference of the particle, so there is no conflict between the
perspectives for potential equations.  The potentials transform in the
same way as the coordinates, so the rotation also affects the vector
potential.

\section{The Li\'{e}nard--Wiechert equations}

The Li\'{e}nard--Wiechert (LW) retardation equations were obtained in
the years 1898 and 1899.  They remain the only known retardation
equations.  The following calculations appear to represent the next term
of the same retardation series, meaning that they are compatible with
the LW equations and their methods of analysis.  The velocity of
conduction electrons in stationary copper wire is so low that the first
term of the series is the only one that is ever needed in those
configurations, so these solutions do not replace the LW solutions in
most applications.  The LW equations are \cite{morse}
\begin{eqnarray}
\Av_\mathrm{LW} &=& q \vv/[r c (1 + \ru \vdot \vv/c)] \\
\psiLW &=& q/[r (1 + \ru \vdot \vv/c)]. \nonumber
\end{eqnarray}
The vector $\rv$ points from the field point to the source, and $\vv$ is
the retarded velocity of the particle.  The electromagnetic fields are
obtained from the retardation solutions with the equations
\begin{eqnarray}
\Ev &=& -\vnabla \psi - (\partial \Av/\partial t)/c \label{efield} \\
\vec H &=& \vnabla \vcross \Av.  \label{hfield}
\end{eqnarray}

The first time derivative of the equations will be needed in the
calculations.  It is convenient to make the substitutions $r=(\rv \vdot
\rv)^{\frac{1}{2}}$ and $\ru=\rv/((\rv \vdot \rv)^{\frac{1}{2}}$ before
performing the differentiations.  After differentiating with respect to the
time at the source, the substitutions $d \rv/d \ts = \vv$ and $d
\vv/d \ts = \av$ are made, where $\av$ is the retarded acceleration.
The solution is simpler if $\rv$ is converted back
to the product of a magnitude and a unit vector as the last step.
The time derivatives become
\begin{flalign}   
d \Av_\mathrm{LW}/d \ts =
q \av /[r c (1 + \ru \vdot \vv/c)]   \label{dlwdts} \\
-q \vv \av \vdot \ru/[r c^2 (1 + \ru \vdot \vv/c)^2] \nonumber \\
-q v^2 \vv/[r^2 c^2 (1 + \ru \vdot \vv/c)^2] \nonumber \\
+q \vv (\ru \vdot \vv)^2/[r^2 c^2 (1 + \ru \vdot \vv/c)^2] \nonumber\\
-q \vv \ru \vdot \vv/[r^2 c (1 + \ru \vdot \vv/c)]  \nonumber
\end{flalign}
\mathspace
\begin{flalign*}
d \psi_\mathrm{LW}/d \ts =
-q \av \vdot \ru/[r c (1 + \ru \vdot \vv/c)^2] \\
-q \ru \vdot \vv/[r^2 (1 + \ru \vdot \vv/c)] \\
-q v^2/[r^2 c (1 + \ru \vdot \vv/c)^2] \\
+q (\ru \vdot \vv)^2/[r^2 c (1 + \ru \vdot \vv/c)^2].
\end{flalign*}

The equations must be parameterized by the time at the field point in
order for them to be usable as retardation equations.  The connection is
$\ts = t - (\rv \vdot \rv)^{\frac{1}{2}}/c$.  After differentiating with
respect to $t$ and selectively substituting $\rv = r \ru$, the equation
becomes $d \ts/d t = 1 - \ru \vdot (d \rv/d t)/c$.  But $d \rv/d t$ is
$d \rv/d \ts~d \ts/d t$, and $d
\rv/d \ts$ is $\vv$, so the solution becomes $d \ts/d t = 1 -
d \ts/d t~ \ru \vdot \vv/c$.  Solving for $d \ts/d t$,
\begin{flalign}
d \ts/d t = 1/(1 + \ru \vdot \vv/c).  \label{dtsdt}
\end{flalign}
$d \Av/d t$ is $d \Av/d \ts~d \ts/d t$, so Eqs
\ref{dlwdts} become
\begin{flalign}   
d \Av_\mathrm{LW}/d t = q \av/[c r (1 + \ru \vdot \vv/c)^2] \label{dlwdt} \\
-q \vv \av \vdot \ru/[c^2 r (1 + \ru \vdot \vv/c)^3]  \nonumber\\
+q \vv (\ru \vdot \vv)^2/[c^2 r^2 (1 + \ru \vdot \vv/c)^3] \nonumber\\
-q \vv v^2/[c^2 r^2 (1 + \ru \vdot \vv/c)^3]  \nonumber \\
-q \vv \ru \vdot \vv/[c r^2 (1 + \ru \vdot \vv/c)^2], \nonumber
\end{flalign}
\mathspace
\begin{flalign*}
d \psi_\mathrm{LW}/d t=-q \av \vdot \ru/[c r (1 + \ru \vdot \vv/c)^3] \\
-q \ru \vdot \vv/[r^2 (1 + \ru \vdot \vv/c)^2]  \\
+q (\ru \vdot \vv)^2/[c r^2 (1 + \ru \vdot \vv/c)^3] \\
-q v^2/[c r^2 (1 + \ru \vdot \vv/c)^3].
\end{flalign*}
The $1/r$ terms are radiative.

\section{\label{calc}{The Thomas terms}}

All but the simplest of retardation problems are intractable if an exact
solution is attempted.  The multivariate Tayor theorm \cite{arfken}
applies to other cases.

One way of performing a multivariate series expansion is with recursive
applications of the Taylor theorem for one variable, but that results in
terms that incomplete in their own order.  The incomplete terms will
occur in subsequent calculations in any case.  Thus if the series
expansion is to order $a^1$ and $v^3$ then the $a^1 v^3$ terms are in
the same order as the $v^4$ terms and must be dropped.  Carrying the
incomplete terms in intermediate calculations is inefficient but
harmless.  It is all right to selectively drop powers of some of the
variables of the full multivariate expansion in the final solution, so
it is also all right to drop the same powers throughout.  When the
equations contain $dt$, it is normally only used for computing the
derivatives, in which case it does not count as one of the variables of
the multivariate expansion.

The Lorentz transform in vector form is
\begin{eqnarray*}
\rv' & = & \gamma (\rv - t \vv) -(\gamma - 1) (\rv - \vv \rv \vdot \vv/v^2) \\
t' & = & \gamma (t - \rv \vdot \vv/c^2),
\end{eqnarray*}
with $\gamma=(1 - v^2/c^2)^{-\frac{1}{2}}$.
When working in series form, the $v^2$ term in the denominator is cancelled
by the $\gamma-1$ term.  That requires that the numerator be initially expanded
to two more powers of velocity than will be needed.  To order $v^3$, the
series solution is
\begin{eqnarray}
\rv' &=& \rv
-t [\vv + v^2 \vv/(2 c^2)]
+\vv \rv \vdot \vv/(2 c^2) \label{lors} \\
t' &=& t [1+ v^2/(2 c^2)]
-(\rv \vdot \vv)/c^2
-v^2 \rv \vdot \vv/(2 c^4). \nonumber
\end{eqnarray}
The calculations will be to order $v^3$ and $a^1$.  There are no $a^2$ terms
in the solution for the first derivative.

The trajectory of the particle is
\begin{eqnarray}
\rones & = & \rv + \vv (\dtsone+\dtstwo) + 1/2~\av(\dtsone + \dtstwo)^2 \label{r1s} \\
\tones & = & \dtsone + \dtstwo - r/c. \nonumber
\end{eqnarray}
The equation is a Taylor expansion of the particle's
position, with the $\av$ term not corresponding accurately to
acceleration in a physical sense when the velocity is high.
The location of the field point is
\begin{eqnarray*}
\ronef & = & 0 \\
\tonef & = & \dtfone + \dtftwo.
\end{eqnarray*}
$(\dtsone+\dtstwo)^2$ expands to $\dtsone^2 + 2 \dtsone \dtstwo + \dtstwo^2$.
The quadratic terms are in the same order as the
$\dtsone~\dtstwo$ term, and they would normally have to be carried.
For this particular calculation, the quadratic terms do not affect the final
solution.  In the interest of brevity, they will not be carried.

The particle must be on the light cone at time $\dtfone$, so $\dtsone$
is not a free parameter.  There is no requirement that it be on the
light cone at time $\dtfone+\dtftwo$, so $\dtftwo$ and $\dtstwo$ can be
set to zero in solving for the light cone condition.  The final solution
would be the same if the light cone constraint were also imposed at time
$\dtfone+\dtftwo$.

The space difference, $\rones - \ronef$, at time $\dtsone$ is $\rv +
\vv~\dtsone$, and the magnitude of the vector is $r + \ru \vdot \vv~\dtsone$.
The time difference, $\tones - \tonef$, is
$-r/c -\dtfone + \dtsone$, leading to the light cone
condition $-r/c -\ru \vdot \vv~\dtsone/c = -r/c + \dtsone-\dtfone$.  The
equation evaluates to $\dtsone = \dtfone/(1+\ru \vdot \vv/c)$.
The solution could have been obtained in a simpler way from Eq \ref{dtsdt}.
A more general approach is needed for other problems, and the method
of successive approximation is usually required.
The solution is expanded in a
series in $v$ then substituted into Eqs \ref{r1s}.  Both ends of the vector are
then transformed to the second frame of reference with the velocity $\vv$.
\begin{flalign*}  
\rtwos = \rv
+r \vv/c
+r \vv \ru \vdot \vv/(2 c^2)
+r v^2 \vv/(2 c^3)  \\
+(\dtfone \dtstwo) [\av
-\av \ru \vdot \vv/c
+\av  (\ru \vdot \vv)^2/c^2 \\
+\vv \av \vdot \vv/(2 c^2)]
\end{flalign*}
\mathspace
\begin{flalign*}
\ttwos = -r/c
-r \ru \vdot \vv/c^2
-r v^2/(2 c^3)
-r v^2 \ru \vdot \vv/(2 c^4) \\
+\dtfone [1
-\ru \vdot \vv/c
-v^2/(2 c^2)
+(\ru \vdot \vv)^2/c^2 \\
+v^2 \ru \vdot \vv/(2 c^3)
-(\ru \vdot \vv)^3/c^3]
+\dtstwo [1 -v^2/(2 c^2)] \\
+(\dtfone \dtstwo) [\av \vdot \vv \ru \vdot \vv/c^3 -\av \vdot \vv/c^2]
\end{flalign*}
\mathspace
\begin{flalign*}
\rtwof = (\dtfone+\dtftwo) [-\vv -v^2 \vv/(2 c^2)]
\end{flalign*}
\mathspace
\begin{flalign*}
\ttwof = (\dtfone+\dtftwo) [1 + v^2/(2 c^2)]
\end{flalign*}
The velocity $\vv_{23} = d\rv/dt$ of the particle in the second frame of
refrence at time $\dtfone$ is needed.  $d\rv$ is obtained by setting
$\dtstwo$ to 0 in the space part of the solution, then subtracting the
result from the full solution.  The result is
\begin{flalign*}  
d\rv = \dtfone \dtstwo
[\av
-\av \ru \vdot \vv/c
+\av (\ru \vdot \vv)^2/c^2
+\vv \av \vdot \vv/(2 c^2)].
\end{flalign*}
Proceeding similarly for the
time part of the solution leads to
\begin{flalign*}   
dt=\dtstwo [1-v^2/(2 c^2)
+\dtfone \av \vdot \vv \ru \vdot \vv/c^3
-\dtfone \av \vdot \vv/c^2].
\end{flalign*}
Then, in series form, $d\rv/dt$ in the second frame of reference becomes
\begin{flalign*}   
\vv_{23}=\dtfone [\av
-\av \ru \vdot \vv/c
+\av v^2/(2 c^2)
+\av (\ru \vdot \vv)^2/c^2  \\
+\vv \av \vdot \vv/(2 c^2)
-\av v^2 \ru \vdot \vv/(2 c^3)].
\end{flalign*}

This velocity is used to transform to the third frame of reference.
$\dtstwo$ was only needed for computing $\vv_{23}$, so it is set to zero
before performing the transform.  The transform is an infinitesimal
transform, so only the first power of velocity in Eqs \ref{lors} is
needed.  The solution is
\begin{flalign*}   
\rthrees =
\rv
+r \vv/c
+r v^2 \vv/(2 c^3)
+r \vv \ru \vdot \vv/(2 c^2) \\
+\dtfone [
+\av r/c
+\av r v^2/c^3
+r \vv \av \vdot \vv/(2 c^3)]
\end{flalign*}
\mathspace
\begin{flalign*}
\tthrees =
-r/c
-r \ru \vdot \vv/c^2
-r v^2/(2 c^3)
-r v^2 \ru \vdot \vv/(2 c^4) \\
+\dtfone [1
-\ru \vdot \vv/c
-v^2/(2 c^2)
+v^2 \ru \vdot \vv/(2 c^3) \\
-(\ru \vdot \vv)^3/c^3
+(\ru \vdot \vv)^2/c^2
-r \av \vdot \ru/c^2
+r \av \vdot \ru \ru \vdot \vv/c^3 \\
-r \av \vdot \vv/c^3
-r v^2 \av \vdot \ru/(2 c^4)
-r \av \vdot \ru (\ru \vdot \vv)^2/c^4]
\end{flalign*}
\mathspace
\begin{flalign*}
\rthreef =
(\dtfone+\dtftwo)[-\vv -v^2 \vv/(2 c^2)]
+\dtfone \dtftwo [   \\
-\av
+\av \ru \vdot \vv/c
-\av v^2/c^2
-\av (\ru \vdot \vv)^2/c^2   \\
-\vv \av \vdot \vv/(2 c^2)]
\end{flalign*}
\mathspace
\begin{flalign*}
\tthreef =
(\dtfone+\dtftwo)[1+ v^2/(2 c^2)]
+\dtfone \dtftwo [\av \vdot \vv/c^2 \\
-\av \vdot \vv \ru \vdot \vv/c^3].
\end{flalign*}

In this frame of reference the potential solution is just the static
Coulomb solution, $\psi=q/r$.  The solution can be viewed as being the
solution for a constant velocity particle moving tangentially to the
trajectory of the accelerated particle.  In its frame of reference the
static potential solution has existed forever, and the distant field
point moves through it.  The motion of the field point does not result
in a vector potential term in its frame of reference.  The past and the
future do not matter for light cone events, so the solution for an
accelerated particle should be the same.  The assumption is subject to
further evaluation.

The scalar solution becomes $\psi = q/[(\rthrees-\rthreef) \vdot
(\rthrees-\rthreef)]^{\frac{1}{2}}$, with $\dtstwo=0$, $\dtftwo=0$.  In
series form,
\begin{flalign}  
\psi = q/r
-q \ru \vdot \vv/(c r)
-q v^2/(2 c^2 r)
+q (\ru \vdot \vv)^2/(c^2 r) \label{pot3} \\
+q v^2 \ru \vdot \vv/(2 c^3 r)
-q (\ru \vdot \vv)^3/(c^3 r)
+\dtfone [-q \ru \vdot \vv/r^2 \nonumber \\
- q \av \vdot \ru/(c r)
+q v^2 \av \vdot \ru/(2 c^3 r)
-6 q \av \vdot \ru (\ru \vdot \vv)^2/(c^3 r) \nonumber \\
+2 q \av \vdot \vv \ru \vdot \vv/(c^3 r)
+7/2 q v^2 \ru \vdot \vv/(c^2 r^2) \nonumber \\
-6 q (\ru \vdot \vv)^3/(c^2 r^2)
+3 q \av \vdot \ru \ru \vdot \vv/(c^2 r)
-q \av \vdot \vv/(c^2 r) \nonumber \\
-q v^2/(c r^2)
+3 q (\ru \vdot \vv)^2/(c r^2)],  \nonumber
\end{flalign}
with $\Av=0$.  The potentials must now be transformed back to the frame
of reference of the field point.
The potentials transform in the same
way as the coordinates.  The velocity of the field point at time
$\dtfone$ in the third frame of refence is obtained in the same way as
in the above calculation for $\vv_{23}$, except that $\dtftwo$ is used
instead of $\dtstwo$.  The solution is
\begin{flalign*}  
\vv_{31} =
-\vv
+\dtfone [
-\av
+\av \ru \vdot \vv/c
-\av v^2/(2 c^2)   \\
-\av (\ru \vdot \vv)^2/c^2
+\vv \av \vdot \vv/(2 c^2) \\
-\av v^2 \ru \vdot \vv/(2 c^3)
-\vv \av \vdot \vv \ru \vdot \vv/c^3].
\end{flalign*}
After transforming the potentials with this velocity,
\begin{flalign}   
\Av = q \vv/(c r)
+q \vv (\ru \vdot \vv)^2/(c^3 r)
-q \vv \ru \vdot \vv/(c^2 r) \label{pot} \\
+q~dt~[\av/(c r)
-2 \av \ru \vdot \vv/(c^2 r)
+\av v^2/(2 c^3 r)    \nonumber \\
-\vv \av \vdot \vv/(2 c^3 r)
+3 \av (\ru \vdot \vv)^2/(c^3 r)
-\vv \av \vdot \ru/(c^2 r)   \nonumber \\
+3 \vv \av \vdot \ru \ru \vdot \vv/(c^3 r)
-v^2 \vv/(c^2 r^2)
-\vv \ru \vdot \vv/(c r^2)   \nonumber \\
+3 \vv (\ru \vdot \vv)^2/(c^2 r^2)]  \nonumber
\end{flalign}
\mathspace
\begin{flalign*}
\psi=
q/r
-q \ru \vdot \vv/(c r)
+q (\ru \vdot \vv)^2/(c^2 r)
-q (\ru \vdot \vv)^3/(c^3 r)  \\
+q~dt~[
-\av \vdot \ru/(c r)
+3 \av \vdot \ru \ru \vdot \vv/(c^2 r)
-6 \av \vdot \ru (\ru \vdot \vv)^2/(c^3 r)   \\
-v^2/(c r^2)
-\ru \vdot \vv/r^2
+3 v^2 \ru \vdot \vv/(c^2 r^2)
+3 (\ru \vdot \vv)^2/(c r^2)   \\
-6 (\ru \vdot \vv)^3/(c^2 r^2)].
\end{flalign*}

The subscript of $\dtfone$ has been dropped in the solution.  The vector
equation has been multiplied by $1/c$ so that the powers of $c$ in the
retardation solutions will be the same as those of the Maxwell
equations.  It is of course possible to work in other systems of units.
It is even possible to use a different system of units for the retarded
potentials than is used for the fields, with the conversion factors
being included in the equations for the fields.

This solution contains both the LW and the Thomas terms, with the full
solution being the simple linear sum of the two.  It will be helpful to
separate the LW and the Thomas terms.  The LW terms could be obtained by
transforming to the frame of reference of the particle at time $t+dt$,
computing the potentials as in Eqs \ref{pot3}, then transforming them
back with the negative of the same velocity.  Another way of obtaining
the same answer is to transform the potentials in Eqs \ref{pot3} back in
two steps, first with the velocity $-\vv_{23}$, then with the velocity
$-\vv$.  It is easier use Eqs \ref{dlwdt} to extrapolate the exact LW
solution to the time $dt$ with the relationship $\Av(dt)=\Av(0)+d \Av/d
t ~dt$, and similarly for the scalar equation.  After converting to
series form,
\begin{flalign}  
\Av_\mathrm{LW}=q \vv/(c r)
-q \vv \ru \vdot \vv/(c^2 r)
+q \vv (\ru \vdot \vv)^2/(c^3 r) \label{lwpot} \\
+dt~[-q \vv \ru \vdot \vv/(c r^2)
+3 q \vv (\ru \vdot \vv)^2/(c^2 r^2) \nonumber\\
-q v^2 \vv/(c^2 r^2)
+\av q/(c r)
-2 \av q \ru \vdot \vv/(c^2 r) \nonumber\\
-q \vv \av \vdot \ru/(c^2 r)
+3 \av q (\ru \vdot \vv)^2/(c^3 r) \nonumber\\
+3 q \vv \av \vdot \ru \ru \vdot \vv/(c^3 r)], \nonumber
\end{flalign}
\mathspace
\begin{flalign*}
\psi_\mathrm{LW}=q/r
-q \ru \vdot \vv/(c r)
+q (\ru \vdot \vv)^2/(c^2 r)
-q (\ru \vdot \vv)^3/(c^3 r) \\
+dt~[-q \av \vdot \ru/(c r)
+3 q \av \vdot \ru \ru \vdot \vv/(c^2 r)
-6 q \av \vdot \ru (\ru \vdot \vv)^2/(c^3 r) \\
-q \ru \vdot \vv/r^2
-q v^2/(c r^2)
+3 q (\ru \vdot \vv)^2/(c r^2)  \\
+3 q v^2 \ru \vdot \vv/(c^2 r^2)
-6 q (\ru \vdot \vv)^3/(c^2 r^2)].
\end{flalign*}
The Thomas terms can now be segregated by subtracting Eqs \ref{lwpot}
from Eqs \ref{pot}.
\begin{eqnarray}
d \Av_\mathrm{T} &=& dt~q \av v^2/(2 c^3 r) -dt~q \vv \av \vdot \vv/(2 c^3 r) \label{sthomas} \\
d \psi_\mathrm{T} &=& 0  \nonumber
\end{eqnarray}
By repeating the calculation with progressively higher powers of
velocity, and observing how the terms evolve, it is not too difficult to
infer the closed form solution.  After dividing through by $dt$, it is
\begin{flalign}
d \Av_\mathrm{T}/d t = q (\av - \vu \av \vdot \vu)(\gamma-1)/[(1 + \ru \vdot \vv/c)^2 c r], \label{dadt}
\end{flalign}
with $\gamma=(1 - v^2/c^2)^{-\frac{1}{2}}$.  Unit velocity vectors are
difficult to work with.  If the expression for $\gamma$ is substituted
and the equation simplified they go away, but the equation is longer.

The solution was validated by expanding it in a series in $\vec a$ and
$\vec v$ then comparing it to the derivation in series form.  The two
calculations were the same to order $v^{30}$, implying that the solution
is exact.  The existence of an exact solution indicates that the entire
derivation could be performed in exact form, but some of the
intermediate expressions are lengthy and difficult to simplify.  The
calculations have not been carried through.

The higher order velocity terms are probably not meaningful without also
carrying the $\adot$ terms, but the compactness of equations that look
like they are exact is nevertheless appealing.

A vector identity can be used to place the solution in the form
\begin{flalign*}
d \Av_\mathrm{T}/d t = q \vu \vcross (\av \vcross \vu) (\gamma-1)/[(1 + \ru \vdot \vv/c)^2 c r],
\end{flalign*}
showing that, as expected, the Thomas terms vanish when the acceleration and
velocity vectors are parallel or anti-parallel.
The full retardation equation is the sum of either of these equations and Eqs \ref{dlwdt}.

The Thomas rotation is of order $a^1 v^1$, but the vector potential is
of order $v^1$, so the Thomas terms in potential form are of order $a^1
v^2$.  When $\av \vdot \vv$ is zero the ratio of the Thomas term in Eq
\ref{sthomas} to the lowest order radiative Maxwell term in Eqs
\ref{dlwdt} is $v^2/(2 c^2)$.  The Thomas terms behave like a
relativistic correction to the Maxwell terms in the far field.

Unless a way can be found to integrate the solution in a general way, it
is necessary to first obtain the solution for the first time derivative,
then integrate it.  The integration is always easy for periodic
solutions, however the static terms are lost.  The static terms could be
retained by directly retarding the $E$ and $B$ fields.

The Thomas terms are difficult to recognize in solutions, so it is often
helpful to multiply the equation by $T$, carry it through the entire
derivation, then set it to 1 in the last step. $T$ sometimes drops out,
meaning that the solution has reduced to the LW result.

The magnitude squared of the total vector potential can be obtained by
multiplying Eq \ref{dadt} by $T$, adding the result to the LW terms in
Eqs \ref{dlwdt}, then computing
\Av \vdot \Av.  To order $v^3$, the result is
\begin{flalign*}   
\Av \vdot \Av = q^2 v^2/(c^2 r^2)
-2 q^2 v^2 \ru \vdot \vv/(c^3 r^2)    \\
+dt~[-2 q^2 v^2 \ru \vdot \vv/(c^2 r^3)
+2 q^2 \av \vdot \vv/(c^2 r^2)   \\
-2 q^2 v^2 \av \vdot \ru/(c^3 r^2)
-6 q^2 \av \vdot \vv \ru \vdot \vv/(c^3 r^2)].
\end{flalign*}
$T$ has dropped out of the solution, showing that the magnitude of the
vector is the same as the LW value.  That means that the full solution
is the LW solution, followed by an infinitesimal space rotation.  Carrying
more powers of velocity in the calculation does not affect the
conclusion.  A space rotation does not affect the scalar potential, so it
is the same as for the LW solution.

When applying the Lorentz transform, a space rotation in the second
frame of reference, no matter how large, leaves the invariant quantity
$r^2-c^2 t^2$ unaltered.  The invariant quantity does not directly apply
to the 4-potential, but since it transforms in the same way as the
coordinates, and since it would be possible to work in a system of units
where the vector potential has the units of distance and the scalar
potential has the units of time, it is likely that mathematical
inconsistencies would arise if the perturbation of the LW vector
potential were more than a space rotation.

The equations look like linear equations, however $a^2$ terms appear in
the second derivatives (not shown), tending to obscure the meaning of
linearity.  Extrapolations around a circle can be built up as a series
of infinitesimal rotations, each of which is linear, but the overall
relationships of a circle are not linear.  As applied to the retardation
equations, each additional infinitesimal rotation requires another
differentiation, so the contravariant tensors can be viewed as achieving
linearization by differentiation, implying that the are not impaired
when compared to nonlinear representations if the rank of the tensor is
sufficient.  The contravariant tensor of the second rank represents the
first derivatives
\cite{gelfand}, which are not very impressive in their capabilities.

The $\adot$ terms drop out of the solution for the first derivative, but
not in the solution for the second derivative.  Other $\adot$ terms
would appear if the retardation solution for the first derivative were
differentiated exclusively in the first frame of reference.  However,
from the perspective of the observer in the other frame of reference,
they would not correctly represent the angular relationships of the
system.  That does not mean that retardation solutions should not be
differentiated in the first frame of reference.  It means that the
derivatives obtained that way are incomplete or inconsistent.  The
argument can be applied recursively, so there is no alternative to using
incomplete or inconsistent retardation equations if they are linear.
(G\"{o}del's proof does not include the qualification to linear
equations.  See www.wikipedia.org.)

Space rotations are the basis of the tensor series, and each tensor to
at least the fifth rank is irreducible \cite{gelfand}, implying that
mathematically complete retardation equations do not exist.  It could be
that real space-time is more degenerate than the tensor series, but
probably not.  However, mathematical degeneration might occur at some
point if the dimensionality of the problem is restricted to four, just
as the $\avdot$ terms are degenerate in the Newton equations.  The
Newton equations represent three copies of one space coordinate and one
time coordinate.  In being vector equations, that is still true in
Minkowski space.  These relationships suggest that when fully extended
to three space dimensions (with the third derivative and the tensor of
the fourth rank), the retardation equations will be degenerate in
$\avdddot$.

Degeneration in $\avdddot$ would imply that the tensor of the fourth
rank plays a special role in the linear relationships of the four
dimensional space, with the equations taking on a new completeness and
consistency in that order.  It is believable that the geometry of the
four dimensional space, linear or otherwise, is of a finite complexity,
and that it can be comprehensively represented.  That would be the end
of the quest unless the dimensionality of real space-time is greater
than four.

The tensor irreducibility theorem only applies to linear equations.  Its
basis is that the multipole order increases with the rank of the tensor,
and multipoles cannot be synthesized from linear combinations of lower
order multipoles.

\section{\label{field}{Some field equations}}
The Proca equations are \cite{morse}
\begin{flalign*}
-\vnabla^2 \Av +\frac{1}{c^2} \frac{\partial^2 \Av}{\partial t^2} = -\vnabla L - \alpha^2 \Av
\end{flalign*}
\moremathspace
\begin{flalign*}
-\vnabla^2 \psi +\frac{1}{c^2} \frac{\partial^2 \psi}{\partial t^2}=\frac{1}{c} \frac{\partial L}{\partial t} -\alpha^2 \psi
\end{flalign*}
\moremathspace
\begin{equation*}
L=\vnabla \vdot \Av + \frac{1}{c} \partial \psi/\partial t
\end{equation*}

The scaler $L$ is known as the Lorentz condition.  After applying the
vector identity $\vnabla^2 \Av = \vnabla(\vnabla \vdot \Av) - \vnabla
\vcross
\vnabla \vcross \Av$, the equations become
\begin{flalign}
     \vnabla\vcross \vnabla\vcross \Av
 - \vnabla(\vnabla \vdot \Av)
     + \frac{1}{c^2}\partial^2 \Av/\partial t^2 = \label{proca2} \\
 -\vnabla L - \alpha^2 \Av  \nonumber
\end{flalign}
\moremathspace
\begin{flalign*}
     -\vnabla^2 \psi + \frac{1}{c^2} \frac{\partial^2 \psi}{\partial t^2}  =
 \frac{1}{c} \frac{\partial L}{\partial t} - \alpha^2 \psi.
\end{flalign*}

The terms on the left are not the Maxwell equations.  However, the
solutions of the LW equations are in the Lorentz guage, and in that
guage $L$ is zero, so when working with those equations the $\vnabla
\vdot \Av$ term in the left part of the vector equation can be replaced
with $-(\partial \psi/\partial t)/c$ Similarly, in the scalar equation
$+(\partial^2 \psi/\partial t^2)/c^2$ can be replaced by
$-(\partial/\partial t) (\vnabla \vdot \Av)/c$.  The equations are now
\begin{flalign*}
     \vnabla\vcross \vnabla\vcross \Av
     +\vnabla(\partial \psi/\partial t)/c
     + \frac{1}{c^2}\partial^2 \Av/\partial t^2 =  -\vnabla L - \alpha^2 \Av
\end{flalign*}
\moremathspace
\begin{flalign*}
     -\vnabla^2 \psi
     -(\partial/\partial t) (\vnabla \vdot \Av)/c =
     \frac{1}{c} \frac{\partial L}{\partial t} - \alpha^2 \psi.
\end{flalign*}

In Eq \ref{proca2}, the $\vnabla(\vnabla \vdot \Av)$ term on the left
cancels the same quantity in $\vnabla L$ when $\vnabla L$ is expanded.
There is no static spherical solution.  Since $L$ is zero, it is all
right to invert its sign, and doing so results in the static solution in
Eq \ref{exp}.  The $\vnabla(\vnabla \vdot \Av)$ term ceased to be a
static term when the substitution $\vnabla \vdot \Av = -(\partial
\psi/\partial t)/c$ was made, so inverting the sign of $L$ in the unmodified
Proca equations does not have the same effect.  The restructured
equations differ from the Proca equations by more than a sign change in
static solutions.

Inverting the sign of $L$ in the scalar equation does not affect the
static scalar solution.  After inverting the sign of $L$ in both
equations, then inverting all the signs of the scalar equation, the
final solution becomes
\begin{flalign}
     \vnabla\vcross \vnabla\vcross \Av
 + \frac{1}{c} \vnabla \frac {\partial \psi}{\partial t}
     + \frac{1}{c^2}\partial^2 \Av/\partial t^2 =
 \vnabla L - \alpha^2 \Av  \label{potvec}
\end{flalign}
\moremathspace
\begin{flalign}
     \vnabla^2 \psi + \frac{1}{c} \frac{\partial}{\partial t} (\vnabla \vdot \Av) =
 \frac{1}{c} \frac{\partial L}{\partial t} + \alpha^2 \psi. \label{potscal}
\end{flalign}

The scalar and vector potentials can be viewed as having the units of
distance in these equations, but with the viewpoint not being unique.

It follows from the method of derivation that these equations are
equivalent to the Proca equations when the solutions are in the Lorentz
gauge, except that the Proca equations do not contain a static vector
solution.  The equations are not equivalent to the Proca equations when
the solutions are not in the Lorentz gauge.  Equations that are similar
to these are derived in the supplemental material at www.s-4.com/tensor,
indicating that the restructuring is not arbitrary.

The terms on the left are the Maxwell equations, and the terms on the
right (with $\alpha=0$) are zero when $L$ is zero.  In this particular
gauge, the equations reduce to the Maxwell equations.  The equations are
therefore fully compatible with the LW equations, even though they are
somewhat more general than the Maxwell equations.

The Lorentz condition is not zero when the solutions contain Thomas
terms.  The $r_0^3$ solutions of the following section are solutions to
these equations (with $\alpha=0$) but they are not solutions to the
Maxwell equations or the Proca equations.  No arguments are presented
that the retardation solutions should be solutions to these equations.
It was simply noticed that they are -- for low order solutions only.

From Eq \ref{hfield}, $\vnabla \vdot \Hv = \vnabla \vdot (\vnabla
\vcross \Av)$.  The right side of the equation is identically zero,
leading to Eq \ref{max1}, which is one of the Maxwell equations.

Reversing the order of differentiation of the second term on the left
side of Eq \ref{potscal} then factoring the left side leads to $\vnabla
\vdot (\vnabla \psi + (\partial \Av/\partial t)/c)$.  From
Eq \ref{efield}, the term can be written as $-\vnabla \vdot \Ev$.
Including the other terms in the scalar equation provides
Eq \ref{max2}.  The equation reduces
to one of the Maxwell equations when $L$ and $\alpha$ are zero.  The
equation does not lead to the lack of charge conservation if the virtual
charge in one region is canceled by virtual charge of the opposite sign
in another region.  It has not been determined if the solutions do
globally conserve charge.

From Eq \ref{efield}, $\vnabla \vcross \Ev = \vnabla \vcross (-\vnabla
\psi - (\partial \Av/\partial t)/c)$. $\vnabla \vcross (\vnabla
\psi)$ is identically zero.  Reversing the order of differentiation of the
$\vnabla \vcross (\partial \Av/\partial t)/c)$ term and substituting
from Eq \ref{hfield} provides Eq \ref{max3}, which is one of the
Maxwell equations.

In Eq \ref{potvec}, $\vnabla \vcross \vnabla \vcross \Av$ is $\vnabla
\vcross \Hv$.  Reversing the order of differentiation in the other two
terms on the left side of the equation and substituting from Eq
\ref{efield} leads to Eq \ref{max4}.  The equation reduces to one of
the Maxwell equations when $L$ and $\alpha$ are zero.

\begin{eqnarray}
\vnabla \vdot \Hv &=& 0 \label{max1} \\
\vnabla \vdot \Ev &=& -(\partial L/\partial t)/c - \alpha^2 \psi \label{max2} \\
\vnabla \vcross \Ev + (\partial \Hv/\partial t)/c &=& 0 \label{max3} \\
\vnabla \vcross \Hv - (\partial \Ev/\partial t)/c &=& \vnabla L - \alpha^2 \Av \label{max4}
\end{eqnarray}

The static solutions could be obtained by directly retarding the $E$ and
$B$ fields.  The scalar $L$ must also be retarded in order to check the
solutions for computational errors.  There is a restriction to low order
solutions. $\alpha$ can usually be set to zero in solutions for the
local region of the cosmos, making it possible to validate the
retardation solutions without knowing the undifferentiated scalar and
vector potentials.

Suppose that $\alpha$ is $1/r_r$, where $r_r$ is the range of the
fields.  The range of the fields is not currently known.  Then, in
spherical coordinates, $\psi = q
\exp(-r/r_r)/r$, $\Av=0$ is a static solution to Eqs \ref{potvec} and \ref{potscal}, with $\nabla^2 \psi =
\alpha^2 \psi$.  When $r \ll r_r$ $\exp(-r/r_r)$ approaches 1 and the
potential solution approaches $\psi=q/r$.
(The units and scaling relationships of $\alpha$ depend on the system of units
utilized.  Alternatively, if the potentials are appropriately scaled with
the units of distance then $\alpha$ is explicitly $1/r_r$, and the
equations shown are of this form.  That entails including a cosmological
constant in local equations.  That may seem wrong at first, but why should
we be different than the rest of the universe?)

Another static solution is
\begin{equation}
A_r = -k^2 r_r \exp(-r/r_r)/r^2 -k^2 \exp(-r/r_r)/r, \label{exp}
\end{equation}
with $\psi=0$, $A_\theta=0$ and $A_\phi=0$.
$k^2$ has the units of distance squared and represents the source strength.
The source strength can also be represented as $b r_r$, which is
more appropriate in a physical sense.  The
Schwarzchild radius is proportional to mass, making it possible
to represent mass with the units of distance, but with the scaling relationships
not directly applying to non-metrical equations.
(This equation is analytically awkward to work with.  Simply substituting
$r=d~r_r$, where $d$ is a number $\ll 1$, is an expedient way of
evaluating its behavior.)

$\vnabla \vcross \Av$ is 0, and $\vnabla^2 \Av$ = $\alpha^2 \Av$.  The
solution is static, so the scalar equation is also satisfied.  When $r
\ll r_r$, $A_r$ is $\approx -k^2 r_r/r^2$ and $\vnabla \vdot
\Av$ is $\approx k^2/(r_r r)$.  The first derivative of the potential
decays with distance at a lower rate than the potential, which is only
possible with exponential equations.  The second derivative, $\vnabla
(\vnabla \vdot
\Av)$, decays approximately as -$k^2/(r_r r^2)$, which represents an inverse
square law vector field in the local region.

The electrostatic and gravitational fields are the only known inverse
square law fields, but it has not been determined if the solutions are
meaningful.

The static scalar and vector solutions define two source terms that
could be retarded, which would have the effect of assimilating
$\alpha=1/r_r$ into the retardation equations.  But $-k^2/(r_r r^2)$
simplifies to $-b/r^2$ if the source strength is written as $b r_r$,
making it possible to scale local solutions without knowing $r_r$.  The
scaling relationships follow from Newtonian gravity in the local region,
without cosmological complications.

The exponential expressions containing $r_r$ must nevertheless be
carried in the differentiations, since $r_r$ does not drop out until
after the first differentiation.  The net effect is that the
cosmological influence, though present, is hidden from us in the
static solutions of this order in the nearby region.

A peek at the solutions for the tensor of the fourth rank can be gained
by simply differentiating again.
\begin{equation*}
\vnabla \vdot (\vnabla (\vnabla \vdot \Av)) = \vnabla \vdot \av =
k^2 \exp(-r/r_r)/(r~rr^3)
\end{equation*}
$\av$ is the acceleration.  $\vnabla \vdot \av$ is zero in Newtonian gravity.

The third derivative of the static scalar solution is
\begin{flalign*}
\vnabla (\vnabla \vdot (\vnabla \psi)) = -\vnabla (\vnabla \vdot \Ev) =
-q \exp (-r/r_r)/(r^2 r_r^2 + r~r_r^3)
\end{flalign*}
As discussed above, the
scalar and vector potentials in these solutions
are in a cosmological system of units.  The MKS scaling relationships
follow from Newtonian gravity, with $k^2$ becoming $r_r G m$.
The MKS electrical scaling relationships
follow from the equation for the $E$ field of a charged particle,
with $q_{\mathrm{mks}} = q/(4 \pi \epsilon_0)$.
The cosmological influence is of first order in the infinitesimal, but
it is much too small to detect in the static solutions for the nearby region.
Its magnitude will need re-evaluation after the dynamic solutions
are obtained.

A term that is first order in the infinitesimal, no matter how small,
cannot be neglected when integrating to cosmological distances.

The unmodified Proca equations contain a similar scalar solution.  In a
quantum context the $\alpha$ of the Proca equations is $m c/\hbar$, and
with the units of $\alpha$ depending on the system of units used.

The differentiations cannot be performed exclusively in the first frame
of reference unless the solution is static.  The Thomas corrections are
required.

There are indications that Eqs \ref{potvec} are only valid for the first
derivatives of the electrical solutions, and it would be surprising if
the same equations are valid for the second derivatives of the radiative
gravitational solutions, even though the static solution is satisfied.
Although undesirable, it is possible to proceed with the development of
retardation equations without yet knowing the associated field
equations.  Matching the retardation equations with field equations will
eventually be essential, but it cannot be done until the radiative
gravitational solutions are obtained.

The radiative solutions to the differential equations will also be
needed, and there could easily be inconsistencies between the two
solutions, inconsistencies that would eventually have to be resolved.
Furthermore, despite the inferences that follow from currently popular
theories, the differential equations of some order might contain the
cosmological redshift, which could require a reformulation of the
retardation problem.

The second derivative gravitational solutions for the local region are
likely to fare better, and they are the appropriate starting point.  But
even in the local region, neglecting the third derivatives causes the
solutions to be approximations.  The relationships are fundamentally
nonlinear, and obtaining exact solutions with linear equations is not
possible.  It remains to be seen if the solutions apply to physical
problems.

Since there are no absolute points of reference, the locally observable
relationships of an acceleration wave do not follow directly from the
retardation solutions.  Two nearby points have to be compared.  The
acceleration wave behaves more like a potential than a field.
Accelerated observers are the only ones that are physically realizable
in the the acceleration wave, so the locally measurable relationships
have to be evaluated in their frame of reference.  Even from the
perspective of a distant observer, Newtonian inferences should not be
applied without further investigation, because the wave does behave more
like a potential than a field, and it is not included in the Newton
equations.  The behavior of the acceleration vector should be Newtonian
when the radiative terms can be neglected.

The radiative solutions of the general theory of relativity are obtained
in a space that is asymptotically flat.  The above relationships cannot
exist in a space where $r_r$ is infinite.  There may be a possibility of
mathematical inconsistencies occurring between the two calculations when
the radius of a radiative solution in spherical coordinates is taken as
being infinite.  Although unintuitive, mathematical singularities at
infinity do occur, and they are capable of causing the form of a
solution to change abruptly in the limit.  The solution obtained by
prematurely setting $r_r = \infty$ in Eq \ref{exp} is not an
approximation.  The derivatives are totally wrong.

Curvature relationships are especially susceptible to odd behavior at
infinity, because the radius of curvature is also infinite in the limit.
It is possible for the calculations to contain unnoticed but undefined
$\infty/\infty$ terms.  In the Newtonian approximation, the acceleration
in the static GR solution is the gradient of the expansion factor.  In
containing nothing more than a quadrupole, the expansion factor is zero
in the radiative far field.  It is important that we know for sure that
it is zero in the far field, as one of the consequences of a non-zero
value would be that a collapsing spherically symmetric mass would
radiate.  Another reason for being sure is that if the GR expansion
factor is in fact both zero and well behaved in the limit then there
will be grave doubts of the validity of the retardation equations.

\section{\label{simpant}{A simple antenna}}

The calculations of this section attempt to integrate around a circle
with the $\av$ terms only.  That can be done perfectly with the Newton
equations, because they are degenerate in $\avdot$.  Retardation
equations are probably not degenerate in $\avdot$, but the problem can
be developed as a series expansion. (The analogy leaves out some steps,
but it may have intuitive merit.)

The calculations are to order $r_0^3$, where $r_0^3$ is the radius of
the orbit for a single charged particle.  The calculation fails in the
$r_0^4$ solution.  The $r_0^4$ terms look like low order terms, and they
are in some sense, but it is also true that the angular velocities are
higher in the near field region.  The Thomas precession is a rotation,
so angular relationships are important.  The particle velocity cannot
exceed $c$, but there is no upper bound to the angular velocities.  That
might cause the neglected $\avdot$ terms to become important in the near
field region.  The same relationship can cause small systems to behave
differently than their larger counterparts.  The LW equations do not
fail in the near field region, but they are insensitive to angular
relationships.

The calculations are straight forward but much to lengthy to show here.
They are shown in the supplemental online material at www.s-4.com/som1.
The material is in the form of raw computer output and it is not very
readable.

After computing the fields, converting the solution back to the
Cartesian system, then setting the x and y coordinates to zero, the far
field Thomas solution along the z axis is
\begin{eqnarray*}
E_\mathrm{x} &=& q r_0^3 \omega^4 \cos (\omega t - \omega z/c)/(2 c^4 z)  \\
E_\mathrm{y} &=& q r_0^3 \omega^4 \sin (\omega t - \omega z/c)/(2 c^4 z)  \\
E_\mathrm{z} &=& 0                                                         \\
H_\mathrm{x} &=& -q r_0^3 \omega^4 \sin (\omega t - \omega z/c)/(2 c^4 z) \\
H_\mathrm{y} &=& q r_0^3 \omega^4 \cos (\omega t - \omega z/c)/(2 c^4 z)  \\
H_\mathrm{z} &=& 0.
\end{eqnarray*}
The orbit is in the x-y plane.
The LW terms in the solution are of the same form, except that
they are multiplied by $r_0 \omega^2/c^2$ rather than $r_0^3 \omega^4/(2 c^4)$.
$r_0 \omega$ is $v$, so the ratio is $v^2/(2 c^2)$.

Both the Thomas and the LW terms represent circularly polarized
Maxwellian radiation.  The Maxwell equations constrain the fields
without specifying what causes them, so all of their relationships are
applicable to unconventional systems when the solutions satisfy the
equations.

Textbooks sometimes attribute the physical basis of solutions to the
Maxwell equations when the actual basis is the LW equations.  The
multipole solutions of field equations are not capable of specifying the
physical properties of the source, so the source terms that the Maxwell
equations can accommodate are more general than most presentations
indicate.  The source terms do not result in multipole terms that are
not already known.  It is rather that the physical basis of the solution
can be unfamiliar.  Unambiguous interpretation of the solutions of field
equations is not possible without retardation equations.  Conversely,
because of the ambiguities of multipole solutions, the Maxwell equations
are more general in a physical sense than was thought at first.  There
are other near field terms in the solution along the z axis that are not
a solution to the Maxwell equations.

The far field Thomas terms along the x axis are
\begin{eqnarray*}
E_\mathrm{x} &=& q r_0^3 \omega^4 \cos (\omega t - \omega x/c)/(2 c^4 x)  \\
E_\mathrm{y} &=& q r_0^3 \omega^4 \sin (\omega t - \omega x/c)/(2 c^4 x)  \\
E_\mathrm{z} &=& 0                                                        \\
H_\mathrm{x} &=& 0                                                        \\
H_\mathrm{y} &=& 0                                                        \\
H_\mathrm{z} &=& q r_0^3 \omega^4 \sin (\omega t - \omega x/c)/(2 c^4 x)  \\
\end{eqnarray*}
The $\Ev$ vector rotates in the x-y plane with a constant magnitude, while
the $\Hv$ vector is parallel to the z axis.  The solution can be
decomposed into two parts.  One of them is equivalent to
an appropriately scaled Maxwellian dipole at the origin and parallel to the y
axis.  After subtracting this component, the residual is
the lone $E_\mathrm{x}$ component,
which is parallel to the direction of propagation.  Even though the other
terms are Thomas terms, only the residual will exhibit any physical behavior
that is not contained in the Maxwell equations.

While the Thomas component of the $E$ field that is parallel to the
propagation direction can be mathematically separated, it is of order
$v^2/c^2$ times the Maxwellian $E$ field in the same solution, and it
probably cannot be physically separated from the Maxwell terms.  With
this interpretation, the component that is parallel to the propagation
direction does not represent a separate form of radiation, but is rather
a minor relativistic correction to the Maxwellian wave.

The gradient of the Lorentz condition defines a third vector in this
solution, but it is not dimensionally consistent with the Thomas
electrical components, so it cannot be combined with any of them in the
same way that the $E$ and $B$ fields are combined to synthesize a
separate form of radiation.  The second rank tensor represents the first
derivatives.  The third rank tensor is more appropriate for the analysis
of the second derivatives of $\vnabla L$.

The tensor decomposition equation \cite{gelfand} was used to obtain the
decomposition products of the third rank tensor.  The calculations were
performed by extrapolating the potentials in space and time to obtain
the second rank tensor.  The decomposition equation is in 3-space, so
the 3+1 space of the first frame of reference is appropriate for the
calculation.  A vector extrapolates as three scalars in 3+1 space.  The
second rank tensor has a 3x3 structure in 3+1 space, and the third rank
tensor is its gradient.  The calculations are shown at
www.s-4.com/tensor.  The calculations were performed in an anisotropic
space of an assumed form.  They could be performed in a more
conventional space, although some interpretation of the behavior of the
$\partial \psi/\partial t$ terms may be necessary in a conventional
space.  The calculations are easily converted to the 4-vector form.

The three vectors and the scalar are
\begin{eqnarray}
\vnabla[\vnabla \vdot \Av + (\partial \psi/\partial t)/c] &=& \vnabla L \label{decomp} \\
(\partial/\partial t) (\vnabla \vcross \Av) &=& \partial \Hv/\partial t \nonumber \\
\vnabla^2 \Av - \vnabla (\partial \psi/\partial t)/c &=&
   -\vnabla \vcross \Hv \nonumber +\vnabla(\vnabla \vdot \Av) \nonumber \\
   & &~~~~ - \vnabla (\partial \psi/\partial t)/c \nonumber \\
\vnabla \vdot [\vnabla \psi + (\partial \Av/\partial t)/c] &=& -\vnabla \vdot \Ev. \nonumber
\end{eqnarray}
The other decomposition products are two quadrupoles and an octupole.
The symmetries of the third rank tensor indicate that its 4-potential
radiative solutions will be much richer than those
of the second rank tensor.  Vector and 4-vector equations are not
necessarily impaired if they are obtained from a tensor of sufficient rank.

As discussed in the online material, calculations in 3+1 space contain
terms that differ by a factor of 3 from their 4-space equivalents.  The
terms are symmetric, and 4-vector equations do not have symmetric terms,
causing the connection to 4-vector equations to be nebulous.  In not
having a direct connection to 4-vector equations, and not affecting the
Maxwell equations, the factor of 3 does not appear in the literature.
The tentative conclusion is that the factor of 3 can be dropped, and it
has been dropped in the decomposition products.  There are indications
that the factor could be carried, and that there is nothing
fundamentally wrong with it, but carrying it would upset many familiar
equations.

After performing two consecutive infinitesimal transforms in the frame
of reference of the particle then integrating the second derivative
twice (not shown), the solution is not a solution to the field
equations.  That is to be expected, since third order field equations
should be required in the next order. (The Thomas precession vanishes if
the $r_0^3$ terms are not carried.  Similarly, the $r_0^4$ terms are
required for a minimal representation of the second derivative.)

Differentiating again in the frame of reference of the particle then
integrating again in the frame of reference of the field point results
in a coupling between the orders.  If the $r_0^4$ terms are dropped in
the second derivative solution then the remaining terms are different
from the solution for the first derivative, but with the only difference
being that all of the Thomas terms are multiplied by 2.  Since the full
solution is the linear sum of the LW and Thomas terms, the solution is
also a solution to the field equations.  The coefficients for the Thomas
terms of $L$ for the $1^{\mathrm{st}}$ through $8^{\mathrm{th}}$
derivatives, after multiple integrations, are approximately 0.5, 1.0,
1.6, 2.1, 2.7, 3.3, 3.9, 4.4.

The series shows that elevating the rank of the associated tensor by one
will have a substantial effect on the magnitude of the Thomas terms.
From a different perspective, the same calculation shows that the
influence of tensors of even very high rank can be folded into the
solution for the first derivative, making it unnecessary to obtain the
complete solution for the higher rank tensor in order to obtain better
accuracy for the first derivative.  It seems that there should be a
better way of obtaining an accurate solution for the first derivative.

It is indicated that the coefficients of the Thomas terms of the above
solutions should be much larger than the values shown, but with the form
of the first derivative solutions not being perturbed by tensors of
higher rank.  It could be that, in a more complete development, the
tensor series that is associated with the retardation series will be
fully orthogonal, with none of the first derivative terms being
perturbed by tensors of higher rank.  The tensor series would be much
better behaved that way, and the possibility is still under
investigation.

\section{\label{lab}{Laboratory evaluation}}

The velocity of conduction electrons in stationary copper wire is so low
that the Thomas terms are undetectable in those configurations.

When the system is mechanically rotating and excited with alternating
current, each conduction electron must be paired with a proton, and the
total charge is enormous.  That causes the magnetic fields due to the
proton and electron currents to be separately enormous.  However, the LW
solution for a rotating current loop is the same as for a stationary
loop when the electrons and protons are retarded separately, so the
enormity of each of the two cancelling fields is not detectable.  The
additional relativistic corrections associated with the Thomas terms do
cause the solution to depend on the mechanical angular velocity.

Even with rotating equipment, the Thomas terms are very weak, and it may
not be feasible to detect them unless a way can be found to separate
them from the Maxwell terms.  It could be that adequate separation is
not achievable for the first derivative solutions.  In that case the
following material may be of interest for higher order solutions.

The solution for a current loop excited by alternating current is
unaltered if the loop is physically rotated, making it impossible for
terms that are first order in the mechanical angular velocity to appear
in the solution, although there are Thomas terms that are quadratic in
velocity.  Discovering a configuration that does have first order terms
would be highly desirable.

Integrating the first time derivative of the retarded potentials is not
well suited to the analysis of the low frequency near field terms in
such solutions.  It will be better to directly retard the $E$ and $B$
fields in order to obtain the static and quasi-static fields.  The
quasi-static near field terms have not yet been investigated, and it may
be worthwhile do so.

As Eqs \ref{sthomas} show, the $\vnabla \psi$ terms makes no
contribution to the Thomas $E$ field.  There are also no $\vnabla
\psi$ terms in the $E$ field near a magnetic toroid that is excited
with alternating current, and a short dipole sensor will not respond to
it, even though it is Maxwellian.  Similar and well understood
experimental complications are to be expected of the non-radiative near
field terms of the Thomas $E$ field.
A Maxwellian detector might
not respond in the expected way unless the term is an approximate
solution to the Maxwell equations.  The complication should not arise
when the solution is predominately just one quasi-static field,
which will frequently be the case in the near field.

As discussed in Section \ref{simpant}, the radiative first derivative
Thomas terms do not appear to be physically separable from the Maxwell
terms, making them undetectable in practice except under extreme
conditions.  The decomposition products of the third rank tensor in Eqs
\ref{decomp} suggest that separation will be easy for the second
dervative solutions.  However, Eq \ref{exp} and its associated equations
indicate that the $\vnabla L$ decomposition product is not not an
electrical field, so it may be better to continue on to the tensor of
the fourth rank, which constrains the relationships among six vectors
\cite{gelfand}.

It is not necessary to know how to build a receiving antenna for
non-Maxwellian radiation in order to detect it.  It is sufficient to
build two transmitting antennas then measure the interaction energy.
The interaction energy can be computed by integrating the sum of the two
fields over a sphere at infinity.  When the antennas are operated at
slightly different frequencies, the interaction energy will be
manifested as a modulation of the current flow in each antenna.  A
similar approach may be useful for near field terms when the available
detectors are not satisfactory.

It is plausible that there are electrical fields associated with an
accelerated or jerked mass, although that will not be known until the
retardation solutions are obtained.  They might provide a readily
accessible desktop method of laboratory evaluation.  Since the field is
the second derivative of the potentials, the terms of experimental
interest may be in the order $\avdot$.  A hammer might be more effective
than a flywheel.

Conduction electrons behave like a fluid with mass when a metal is
accelerated, causing the metal to acquire a weak electrostatic dipole
moment.  Abrupt changes in the acceleration are probably capable of
producing a magnetic field.  In the Barnett effect \cite{barnett}, a
rotating mass acquires a weak and sustained magnetic field, which is
attributed to the spin of the electron.  Weak piezoelectric effects may
also need consideration in some configurations.  It will obviously be
important to not confuse known effects with the retardation solutions.

Inferring the electrical scaling relationships of gravitational
solutions will require some plausible guesses, since the two fields are
perfectly orthogonal in lower order representations.  The connection
between the $b r_r$ source term in Eq \ref{exp} and the Schwarzchild
radius, when considered in relation to to the Dirac field limits,
provides a plausible method of inferring the scaling.  The connection is
that the Schwarzchild radius and the Dirac limits both represent
limiting conditions, and the electrical and gravitational fields are
probably on an equal footing in the limit.  The calculation requires a
pseudo-metrical interpretation of the retardation equations.  The Dirac
field limits are not sharply defined, and a pseudo-metrical
approximation will not be very accurate, but order of magnitude
estimates are adequate for designing laboratory equipment.  The vector
potential points inward in the gravitational solution.  When represented
as a displacement in space, it cannot have a magnitude greater than the
radius of the mass, which can be the basis of a rough scaling
relationship.  These relationships are developed further in
Ref.~\onlinecite{osborn}.  The material at www.s-4.com/pulsar is more
current and more complete.

The scaling relationships of the electrical solutions of the general
theory of relativity have no observational confirmation and they are not
commensurate with the Dirac field limits, so they will not be useful in
developing the coupling between the electrostatic and gravitational
fields in dynamic systems.  There is probably no coupling in static
systems, just as there is no coupling between the $E$ and $B$ fields in
static solutions.  Classical calculations cannot be commensurate with
quantum results until the Planck constant is assimilated.

But by of the identity $\alpha = \mu_0 c q_e^2/(2 h)$ (MKS units),
assimilating the fine structure constant is equivalent to assimilating
the Planck constant.  The fine structure constant is a pure number
quantity.  We need a clear and precise understanding of why the geometry
of the four dimensional space contains a ratio that is this particular
value.  Until we do, $\alpha$ can be used empirically, along with the
retardation equations, in developing the scaling relationships of the
coupling of the electrostatic and gravitational fields.  The empirical
relationships will eventually be useful in inferring the essential
meaning of the constant.  The numerical value of the constant should
then be computable.  In being purely geometrical, the solution should be
exact
\textit{if} the dimensionality of real space-time is four.

It has been established by the equivalence principle and its
consequences that acceleration and gravity are not distinguishable with
the second derivative.  It is therefore possible for a physical constant
that was originally discovered in an electrical context to have a
meaning that cannot be inferred in an electrical context.

There are pitfalls in relying on empirical geometry, with the worst of
them being a loss of meaning rather than a loss of numerical accuracy.

\end{document}